\documentclass[letter,twocolumn]{jpsj3} 
\usepackage{bm,amssymb,amsmath}
\usepackage{graphicx}   
\newcommand{\simg}{\stackrel{>}{_\sim}}
\newcommand{\siml}{\stackrel{<}{_\sim}}

\title{
\begin{center}
{\large {\bf 
Dynamical Mean-Field Study of Local Pairing Interaction Mediated by Spin and Orbital Fluctuations in Iron Pnictide Superconductors
}}
\end{center}}
\author{Takemi {\sc Yamada}\thanks{E-mail address: takemi@phys.sc.niigata-u.ac.jp} and Yoshiaki {\sc \=Ono}}

\inst{Department of Physics, Niigata University, Ikarashi, Nishi-ku, Niigata, 950-2181, Japan}

\abst{
We investigate the two-orbital Hubbard model, which reproduces the electron and hole Fermi surfaces in the iron pnictide superconductors, in the presence of the Jahn-Teller electron-phonon coupling by using the dynamical mean-field theory. When the intra- and inter-orbital Coulomb interactions, $U$ and $U'$, increase with $U=U'$, both the local spin and orbital susceptibilities, $\chi_{s}$ and $\chi_{o}$, increase with $\chi_{s}=\chi_{o}$ because of the spin-orbital symmetry. Due to the Hund's rule coupling $J$, $\chi_{s}$ is enhanced and dominates over $\chi_{o}$ resulting in the repulsive local pairing interaction $V_{\rm loc}>0$, while due to the electron-phonon coupling $g$, $\chi_{o}$ is enhanced and dominates over $\chi_{s}$ resulting in the attractive one $V_{\rm loc}<0$ which induces the intra-orbital $s$-wave pairing. Remarkably, $V_{\rm loc}$ is weakly dependent on doping and can be attractive for heavily electron-doped regime where the superconductivity is observed without Fermi surface nesting. 
}
\kword{iron pnictide superconductor, multi-orbital system, Jahn-Teller electron-phonon coupling, orbital fluctuation, magnetic fluctuation, dynamical mean-field theory}

\begin{document}
\maketitle

The iron pnictide superconductors exhibit the common feature of phase diagrams, where parent compounds show the tetragonal-orthorhombic structural transition and the stripe-type antiferromagnetic (AFM) transition both of which are suppressed by carrier doping $x$ resulting in the high-$T_c$ superconductivity\cite{JACS.130.3296,AdvPhys.59.803}. 
When approaching the AFM transition, the AFM fluctuation observed by the NMR experiments\cite{PhysRevLett.104.037001} is found to be enhanced, while, when approaching the structural transition, the $d_{xz}$-$d_{yz}$ ferro-orbital (FO) fluctuation (or the $O_{x^2 -y^2}$ ferroquadrupole fluctuation)\cite{notation} responsible for the softening of the elastic constant $C_{66}$ observed by the ultrasonic experiments\cite{JPSJ.81.024604,JPSJ.80.073702} is found to be enhanced. 
Correspondingly, two distinct $s$-wave pairings: the $s_{\pm}$-wave with sign change of the order parameter between the hole and the electron Fermi surfaces (FSs) mediated by the AFM fluctuation\cite{PhysRevLett.101.057003,PhysRevLett.101.087004} and the $s_{++}$-wave without the sign change mediated by the FO fluctuation\cite{JPSJ.79.123707,SSC.152.701} and by the antiferro-orbital (AFO) fluctuation\cite{PhysRevLett.103.177001} which is also responsible for the softening of $C_{66}$ through the two-orbiton process\cite{PhysRevB.84.024528}, were proposed.

Recent experiments have revealed that the high-$T_c$ superconductivity is realized even in the case with heavily electron-doped compounds such as RFeAsO$_{1-x}$H$_{x}$ (R=Sm, Ce, La)\cite{PhysRevB.84.024521,PhysRevB.85.014514,NatureComm.3.943} up to $x\sim 0.5$ and A$_x$Fe$_2$Se$_2$ (A=K, Cs, Rb)\cite{PhysRevB.82.180520,NatureMat.10.273} where the large electron FSs are observed without the hole FSs. 
In this case, the mechanisms based on the AFM\cite{PhysRevLett.101.057003,PhysRevLett.101.087004} and the AFO\cite{PhysRevLett.103.177001} fluctuations, which are enhanced due to the nesting between the electron and hole FSs, seem to be insufficient for explaining the superconductivity. As for the mechanism based on the FO fluctuation\cite{JPSJ.79.123707} which is enhanced due to the coupling between the $d_{xz}$-$d_{yz}$ orbital fluctuation and the orthorhombic phonon, the superconductivity does not need the FS nesting effect but is restricted near the tetragonal-orthorhombic structural transition with small $x$ within the random phase approximation (RPA).

Generally, in the magnetic fluctuation mechanism, the the pairing interaction $V(\bm{q})$ with wavevector $\bm{q}$ is repulsive and then the strong $\bm{q}$-dependence of $V(\bm{q})$ realized near the magnetic ordered phase is crucial for the superconductivity. On the other hand, in the orbital fluctuation mechanism, $V(\bm{q})$ is attractive and then the strong $\bm{q}$-dependence of $V(\bm{q})$ realized near the orbital ordered phase is not necessary for the superconductivity. When the local component of the orbital fluctuation is relatively larger than that of the magnetic fluctuation, the local component of the pairing interaction $V_{\rm loc}$, which is nothing but the $\bm{q}$-averaged value of $V(\bm{q})$, becomes attractive and induces the $s_{++}$-wave pairing, even far away from the ordered phases.

In this letter, we investigate the local paring interaction $V_{\rm loc}$ mediated by the spin and orbital fluctuations by using the dynamical mean-field theory (DMFT)\cite{RevModPhys.68.13} which becomes exact in infinite dimensions ($d=\infty$) and enables us to sufficiently include the local correlation effect beyond the RPA. We employ the two-orbital Hubbard model\cite{PhysRevB.77.220503,JPSJ.78.083704}, which reproduces the electron and hole FSs in the iron pnictides, in the presence of the coupling between the orbital fluctuation and the Jahn-Teller (JT) phonon corresponding to the orthorhombic mode responsible for the softening of $C_{66}$ as discussed in the previous work\cite{JPSJ.79.123707,SSC.152.701}. Although the present model is a simplified version of the multi-orbital electron-phonon models for the iron pnictides\cite{JPSJ.79.123707,SSC.152.701,PhysRevLett.103.177001,PhysRevB.84.024528,PhysRevB.82.064518}, the essential feature of the local magnetic and orbital fluctuations, which is crucial for determining $V_{\rm loc}$, is expected to be well described.

Our model Hamiltonian is given by 
\begin{align}
H=H_{\rm 0}+H_{\rm int}+H_{\rm ph}+H_{\rm el-ph} 
\label{eq:H} 
\end{align}
with the kinetic part of the Hamiltonian: 
\begin{align}
 &H_{\rm 0}=\!\!\sum_{\bm{k}\sigma}
\left(d_{\bm{k}1\sigma}^{\dagger}~\!d_{\bm{k}2\sigma}^{\dagger}\right)
\!\hat{H_{\bm{k}}}\!
\left(\begin{array}{l}
\!\!\!d_{\bm{k}1\sigma}\!\!\!\!\\
\!\!\!d_{\bm{k}2\sigma}\!\!\!\!\\
\end{array}\right)\!, \ 
\hat{H_{\bm{k}}}\!=\!\left(\begin{array}{ll}
\!\!\!\!\epsilon_{\bm{k}1} \!\!\! &\!\!\!\epsilon_{\bm{k}12}\!\!\!\! \\
\!\!\!\!\epsilon_{\bm{k}12}\!\!\! &\!\!\!\epsilon_{\bm{k}2}\!\!\!\! \\
\end{array}\right)\!
\label{eq:H0}
\end{align}
where $d_{\bm{k}l\sigma}$ is the annihilation operator for a Fe $3d$ electron with the wavevector $\bm{k}$ and  the spin $\sigma$ in the orbital $l=1,2(=d_{xz},d_{yz})$, and the energies $\epsilon_{\bm{k}l}$ and $\epsilon_{\bm{k}12}$ are determined so as to reproduce the electron and hole FSs in the iron pnictides\cite{PhysRevB.77.220503}. The Coulomb interaction part $H_{\rm int}$ includes the intra- and inter-orbital direct terms $U$ and $U'$, the Hund's rule coupling $J$ and the pair transfer $J'$. For simplicity, we assume the relations $U=U'+2J$ and $J=J'$ which are satisfied in the isolated atom but not generally in the crystal\cite{JPSJ.77.123701,PhysRevB.81.054518}. 
The phonon and the electron-phonon interaction parts are given by 
\begin{align}
&H_{\rm ph}+H_{\rm el-ph}=\sum_{i}\omega_{0}b_{i}^{\dagger}b_{i}+g\sum_{i}\left(b_{i}+b_{i}^{\dagger}\right)\tau_{zi},
\end{align}
where $b_{i}$ is the annihilation operator for a JT phonon at site $i$ with the frequency $\omega_{0}$, which is coupled to the longitudinal orbital fluctuation, $\tau_{zi}=\sum_{\sigma}(n_{i1\sigma}-n_{i2\sigma})$ with $n_{il\sigma}=d_{il\sigma}^{\dagger}d_{il\sigma}$, through the electron-phonon coupling $g$.

To solve the model eq.(\ref{eq:H}), we use the DMFT\cite{RevModPhys.68.13} in which the lattice model is mapped onto an impurity Anderson model embedded in an effective medium which is determined so as to satisfy the self-consistency condition 
\begin{align}
\left[{\cal \hat{G}}^{-1}(z)-\hat{\Sigma}(z)\right]^{-1}
=\frac{1}{N}\sum_{\bm{k}}\left[z-\hat{H}_{\bm{k}}-\hat{\Sigma}(z)
\right]^{-1}, 
\label{eq:sce}
\end{align}
where $\hat{\Sigma}(z)$ and ${\cal\hat{G}}(z)$ are the $2\times2$ matrix representations of the impurity (local) self-energy and the bare impurity Green's function, respectively, and $z$ is the complex frequency. 

We write the effective pairing interaction for the spin-singlet state mediated by the spin and charge-orbital fluctuations in the $4\times4$ matrix representation as\cite{PhysRevB.69.104504}
\begin{eqnarray}
\hat{V}(\bm{q},z)\!\!\!\!&=&\!\!\!\!
 \frac{3}{2}\hat{\Gamma}_{s}(z)\hat{\chi}_{s}(\bm{q},z)\hat{\Gamma}_{s}(z)
-\frac{1}{2}\hat{\Gamma}_{c}(z)\hat{\chi}_{c}(\bm{q},z)\hat{\Gamma}_{c}(z)
 \nonumber \\   \!\!\!\!&+&\!\!\!\!
 \frac{1}{2}(\hat{\Gamma}_{s}^{(0)}+\hat{\Gamma}_{c}^{(0)}),
\label{eq:pair}
\end{eqnarray}
where $\hat{\Gamma}_{s(c)}(z)$ is the local irreducible spin (charge-orbital) vertex in which only the external frequency ($z$) dependence is considered as a simplified approximation.\cite{zdependence} The corresponding bare vertex is given by: 
$[\Gamma_{s(c)}^{(0)}]_{llll}=U(U-2g^2 D(z))$, 
$[\Gamma_{s(c)}^{(0)}]_{ll'll'}=U'(-U'+2J)$, 
$[\Gamma_{s(c)}^{(0)}]_{lll'l'}=J(2U'-J+2g^2 D(z))$ and 
$[\Gamma_{s(c)}^{(0)}]_{ll'l'l}=J'(J')$ 
with the bare phonon Green's function $D(z)=2\omega_0/(\omega_0^2-z^2)$, 
where $l'\neq l$ and the other matrix elements are 0\cite{JPSJ.77.123701,PhysRevB.81.054518}. 
Within the DMFT, the spin (charge-orbital) susceptibility can be expressed as 
$\hat{\chi}_{s(c)}(\bm{q},z)=
[\hat{\chi}_0(\bm{q},z)^{-1} -(+)\hat{\Gamma}_{s(c)}(z)]^{-1}$, 
where 
$\hat{\chi}_0(\bm{q},i\omega)=-T\sum_{\bm{k}\epsilon}
\hat{G}(\bm{k}+\bm{q},i\epsilon+i\omega)\hat{G}(\bm{k},i\epsilon)$ 
with the lattice Green's function 
$\hat{G}(\bm{k},z)=[z-\hat{H}_{\bm{k}}-\hat{\Sigma}(z)]^{-1}$, 
and the local vertex can be expressed as 
$\hat{\Gamma}_{s(c)}(z)=\hat{\chi}_{0}^{-1}(z)-\hat{\chi}_{s(c)}^{-1}(z)$, 
where 
$\hat{\chi}_0(i\omega)=-T\sum_{\epsilon}
\hat{G}(i\epsilon+i\omega)\hat{G}(i\epsilon)$ 
with the local (impurity) Green's function 
$\hat{G}(z)=[{\cal \hat{G}}^{-1}(z)-\hat{\Sigma}(z)]^{-1}$. 
When we replace $\hat{\Gamma}_{s(c)}(z)$ with $\hat{\Gamma}_{s}^{(0)}$  and neglect $\hat{\Sigma}(z)$, eq. (\ref{eq:pair}) yields the RPA result of $\hat{V}(\bm{q},z)$\cite{PhysRevB.82.064518,JPSJ.77.123701,PhysRevB.81.054518,PhysRevB.69.104504}. Therefore, eq.(\ref{eq:pair}) is a straightforward extension of the RPA result of $\hat{V}(\bm{q},z)$ to include the vertex and the self-energy corresctions within the DMFT without double counting.

The superconductivity can be examined by solving the Eliashberg equation with $\hat{V}(\bm{q},z)$ given in eq.(\ref{eq:pair}) where the 1st and 2nd terms of r.h.s. yield $\bm{k}$-dependent anomalous self-energies (or gap functions) which includes $1/d$ corrections for the DMFT\cite{RevModPhys.68.13}. In this letter, we focus only on the intra-orbital part of the local paring interaction $V_{\rm loc}\equiv[\hat{V}(z=0)]_{llll}$ with 
$\hat{V}(z)=\frac{1}{N}\sum_{\bm{q}}\hat{V}(\bm{q},z)=
\frac{3}{2}\hat{\Gamma}_{s}(z)\hat{\chi}_{s}(z)\hat{\Gamma}_{s}(z)-\frac{1}{2}\hat{\Gamma}_{c}(z)\hat{\chi}_{c}(z)\hat{\Gamma}_{c}(z)+\frac{1}{2}(\hat{\Gamma}_{s}^{(0)}+\hat{\Gamma}_{c}^{(0)})$, 
as it is the most dominant contribution for the $s_{++}$-wave pairing due to the orbital fluctuation\cite{JPSJ.79.123707,SSC.152.701}.

In the actual calculations with the DMFT, we solve the effective impurity Anderson model coupled to the JT phonon by using the exact diagonalization (ED) method for a finite-size cluster to obtain the local quantities such as $\hat{\Sigma}(z)$ and $\hat{\chi}_{s(c)}(z)$ at $T=0$, where we set the site number $N_s=4-6$ and the cutoff of the phonon number $N_{b}=20$\cite{PhysC.426.330,JPSJ.79.054707}. The tight-binding parameters are set to be the same in Ref.\cite{PhysRevB.77.220503} where the total band width is $W=12$, and we set the phonon frequency $\omega_{0}=0.01W$.  Using the ED method, we calculate the several physical quatities as follows: the renormalization factor corresponding to the inverse effective mass $Z=(1-\frac{\rm{d} \Sigma(z)}{\rm{d} z}|_{_{z=0}})^{-1}=(m^*/m)^{-1}$, 
the local charge, spin and orbital fluctuations 
$\langle\delta n^{2}\rangle$ with $\delta n=n-\langle n\rangle$, 
$\langle\bm{S}^{2}\rangle$ with $\bm{S}=\frac{1}{2}\sum_{l}\sum_{\alpha\beta}d_{il\alpha}^{\dagger}\bm{\sigma}_{\alpha\beta}d_{il\beta}$ and 
$\langle\tau_{z}^{2}\rangle$, 
and the local spin and orbital susceptibilities 
$\chi_{s}=4\langle \langle S_{z}|S_{z} \rangle \rangle|_{z=0}$ and 
$\chi_{o}=\langle \langle \tau_{z}|\tau_{z} \rangle \rangle|_{z=0}$, 
together with the intra-orbital local paring interaction $V_{\rm loc}$ mentioned above. 

\begin{figure}[t]
\begin{center}
\includegraphics[width=6.50cm]{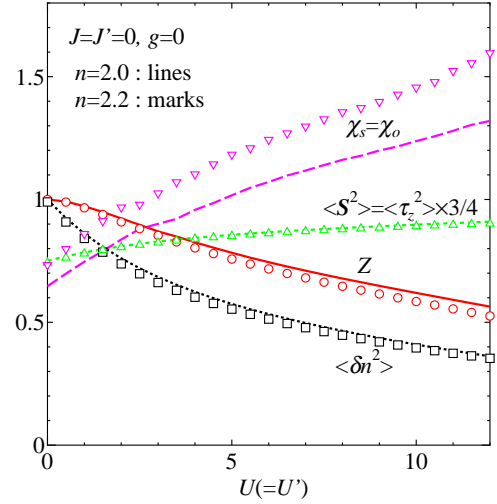}
\caption{(Color online) 
$U(=U')$ dependence of $Z$, $\langle\bm{S}^{2}\rangle$, $\langle\delta n^{2}\rangle$, $\chi_{s}$ and $\chi_{o}$ with $J=J'=g=0$ 
for $n=2$ (lines) and $n=2.2$ (marks).}
\label{fig:Fig1}
\end{center}
\end{figure}


Fig. \ref{fig:Fig1} shows the several physical quantities mentioned above as functions of $U(=U')$ with $J=J'=g=0$ at half-filling $n=2$ and away from half-filling $n=2.2$.  When the electron correlation increases with $U=U'$, $Z$ and $\langle \delta n^{2}\rangle$ decrease while $\langle\bm{S}^{2}\rangle$ and $\langle\tau_{z}^{2}\rangle$ increase with $\langle\bm{S}^{2}\rangle=\frac{3}{4}\langle\tau_{z}^{2}\rangle$ as the double-occupancy probabilities take the same value: $\langle n_{l\uparrow}n_{l\downarrow}\rangle=\langle n_{l\uparrow}n_{l'\downarrow}\rangle=\langle n_{l\uparrow}n_{l'\uparrow}\rangle$ with $l\neq l'$ because of the spin-orbital symmetry\cite{JPSJ.79.054707}. Correspondingly, $\chi_{s}$ and $\chi_{o}$ increase with $\chi_{s}=\chi_{o}$ while the charge susceptibility decreases (not shown) with increasing $U$. For $n=2$, we also observe the Mott metal-insulator transition at a critical interaction $U_c(=U'_c)\sim 2.5W$, where $Z=0$  for $U>U_c$, while when $U \to U_c$ for $U<U_c$, $Z\to 0$ and $\chi_{s}=\chi_{o}\to \infty$ (not shown), as previouly observed in the multi-orbital Hubbard model\cite{PhysRevB67.035119}. For $n=2.2$, the $U$ dependence of the physical quantities is almost similar to that for $n=2$ as shown in Fig. \ref{fig:Fig1}, except for the Mott transition which is observed exclusively for integer fillings\cite{PhysRevB67.035119}. We note that, althogh the $\bm{q}$ dependence of $\hat{\chi}_{s}(\bm{q},z)$  largely dependes on doping responsible for the Fermi surface nesting, the $\bm{q}$-averaged value, i. e., the local susceptibility $\hat{\chi}_{s}(z)$ is weakly dependent on doping.

\begin{figure}[t]
\begin{center}
\includegraphics[width=6.50cm]{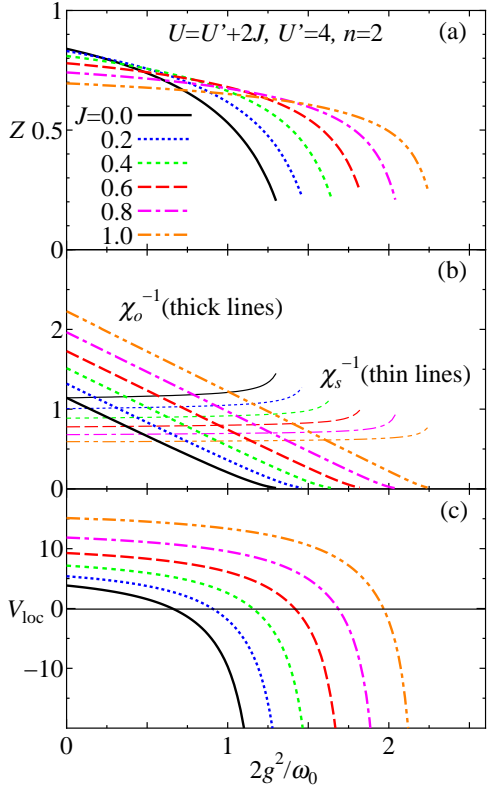}
\caption{(Color online) 
$2g^{2}/\omega_{0}$ dependence of $Z$ (a), $\chi_{s(o)}^{-1}$ (b) and $V_{\rm loc}$ (c) for several values of $J(=J')$ with $U'=4$ and $n=2$. 
}\label{fig:Fig2}
\end{center}
\end{figure}

\begin{figure}[t]
\begin{center}
\includegraphics[width=6.50cm]{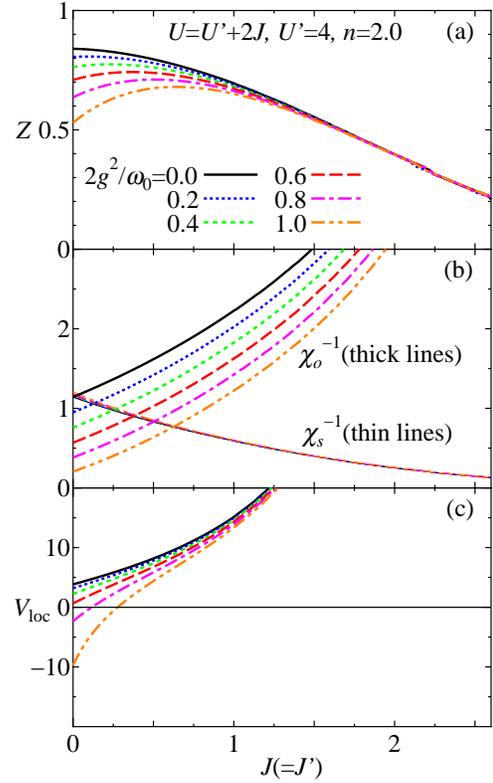}
\caption{(Color online) 
$J$ dependence of $Z$ (a), $\chi_{s}$ and $\chi_{o}$ (b) and $V_{\rm loc}$ (c) 
for several values of $2g^{2}/\omega_{0}$ with $U'=4$ and $n=2$. 
}\label{fig:Fig3}
\end{center}
\end{figure}

Next, we consider the effects of the Hund's rule coupling $J(=J')$ and the JT electron-phonon coupling $g$. In Figs. \ref{fig:Fig2} (a)-(c), $Z$, $\chi_{s(o)}^{-1}$ and $V_{\rm loc}$ are plotted as functions of $2g^{2}/\omega_{0}$ for several values of $J$ with $U'=4$ and $n=2$. When $2g^{2}/\omega_{0}$ increases, $Z$ decreases with increasing $\chi_{o}$ due to the strong orbital-lattice coupling effect, while $\chi_{s}$ slightly decreases. Correspondingly, $V_{\rm loc}$ decreases with increasing $2g^{2}/\omega_{0}$ and finally becomes negative where the attractive term due to $\chi_{o}$ dominates over the repulsive term due to $\chi_{s}$ (see eq.(\ref{eq:pair})). Then, the intra-orbital $s$-wave ($s_{++}$-wave) pairing is expected to be realized in the intermediate coupling regime where $V_{\rm loc}<0$ and $Z=1/2 \sim 1/3$. In the strong coupling regime, we also observe the bipolaronic transition at a critical coupling $g_c$ where $Z\to 0$ with  $g\to g_c$ together with $\chi_{o}\to \infty$ and $V_{\rm loc}\to \infty$, although it is difficult to obtain a fully convergent solution with $Z\sim 0$. The effect of $J$ enhances $\chi_{s}$ while suppresses $\chi_{o}$ (see also Fig. \ref{fig:Fig3}).


Figs. \ref{fig:Fig3} (a)-(c) show the $J$ dependence of $Z$, $\chi_{s(o)}^{-1}$ and $V_{\rm loc}$ for several values of $2g^{2}/\omega_{0}$ with $U'=4$ and $n=2$. When $J$ increases, $Z$ monotonically decreases for $2g^{2}/\omega_{0}=0$ while it shows a maximum at $J\sim 2g^{2}/\omega_{0}$ for $2g^{2}/\omega_{0}\ne 0$. $\chi_{s}(\chi_{o})$ increases (decreases) with increasing $J$ resulting in a crossover between the following two regimes: $J\siml 2g^{2}/\omega_{0}$ with $\chi_{s}<\chi_{o}$ and $J\simg 2g^{2}/\omega_{0}$ with $\chi_{s}>\chi_{o}$. 
Then, the large effective mass $m^*/m=Z^{-1}\gg 1$ is observed in the two distinct regimes with $J\ll 2g^{2}/\omega_{0}$ ($J\gg 2g^{2}/\omega_{0}$) where $\chi_{o}(\chi_{s})$ dominates over $\chi_{s}(\chi_{o})$ due to the strong coupling (correlation) effect, while the moderate effective mass $m^*/m=Z^{-1}=2 \sim 3$ is observed in the intermediate regime with $J\sim 2g^{2}/\omega_{0}$ where $\chi_{s}$ and $\chi_{o}$, both of which are largely enhanced by $U(=U')$ as shown in Fig. \ref{fig:Fig1}, compete to each other resulting in a maximum of $Z$ as a fully non-perturbative effect. This intermediate regime with $\chi_{o}$ being a little larger than $\chi_{s}$ is responsible for the $s_{++}$-wave pairing with $V_{\rm loc}<0$ and seems to be consistent with the iron-pnictide superconductors where both the spin and orbital fluctuations are large while the renormalization of the band width is moderate $Z=1/2 \sim 1/3$\cite{AdvPhys.59.803}. 

\begin{figure}[t]
\begin{center}
\includegraphics[width=6.50cm]{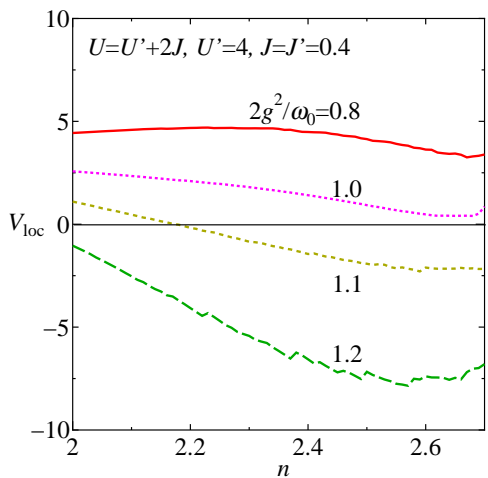}
\caption{(Color online) 
$n$ dependence of  $V_{\rm loc}$ for several values of $2g^{2}/\omega_{0}$ with $U'=4$ and $J=J'=0.4$. 
}\label{fig:Fig4}
\end{center}
\end{figure}

Finally, we study the doping dependence of $V_{\rm loc}$. 
In Fig. \ref{fig:Fig4}, we plot  $V_{\rm loc}$ as a function of $n$ for several values of $2g^{2}/\omega_{0}$ with $U'=4$ and $J=0.4$. For small $2g^{2}/\omega_{0}$, $V_{\rm loc}$ is repulsive as $\chi_{s}$ dominates over $\chi_{o}$. When $2g^{2}/\omega_{0}$ increases, $\chi_{o}$ increases and then dominates over $\chi_{s}$ resulting in $V_{\rm loc}<0$ as shown in Fig. \ref{fig:Fig2} (c) for $n=2$. Remarkably, $V_{\rm loc}$ is weakly dependent on $n$ and can be attractive even far away from half-filling, where the $s_{++}$-wave superconductivity is expected to be realized without the nesting between the electron and hole FSs as recently observed in the heavily electron-doped compounds\cite{PhysRevB.84.024521,PhysRevB.85.014514,NatureComm.3.943,PhysRevB.82.180520,NatureMat.10.273}.

In summary, we have investigated the local pairing interaction $V_{\rm loc}$ mediated by the spin and orbital fluctuations in the two-orbital Hubbard model for iron pnictides by using the DMFT+ED method which enables us to sufficiently include the local correlation effects due to the Coulomb and JT electron-phonon interactions beyond the RPA. It has been found that $V_{\rm loc}$ is weakly dependent on doping and becomes attractive in the intermediate regime with $J\siml 2g^{2}/\omega_{0}$ where the local orbital susceptibility is a little larger than the local spin susceptibility and the band width renormalization is moderate $1/2 \sim 1/3$. The result is responsible for the $s_{++}$-wave pairing even far away from the half-filling without Fermi surface nesting and seems to be consistent with the high-$T_c$ superconductivity observed in the heavily electron-doped compounds\cite{PhysRevB.84.024521,PhysRevB.85.014514,NatureComm.3.943,PhysRevB.82.180520,NatureMat.10.273}. 

Based on the present study, we can calculate the $\bm{q}$-dependent spin (charge-orbital) susceptibility $\hat{\chi}_{s(c)}(\bm{q},z)$ and determine the instability towards the ordered phase when the corresponding susceptibility diverges. Preliminary results shows that the stripe-type AFM observed for $J\simg 2g^{2}/\omega_{0}$ is largely suppressed as compared to the RPA result due to the strong correlation effect within the DMFT while the suppression of the FO observed for $J\siml 2g^{2}/\omega_{0}$ is not so large. We also obtain some results for the superconductivity by solving the Eliashberg equation and find that the $s_{\pm}$-wave occurs near the AFM while the $s_{++}$-waves occur near the FO, where the $s_{++}$-region is qualitatively consistent with the $V_{\rm loc}<0$ region shown in this letter but expands to a rather smaller value of $2g^{2}/\omega_{0}$. Detailed calculations for the present model, together with those for a more realistic 5-orbital model, are now under way.

\section*{Acknowledgments}
The authors thank Y. Yanagi for useful comments and discussions. 
This work was partially supported by a Grant-in-Aid for Scientific Research from the Ministry of Education, Culture, Sports, Science and Technology, and also by a Grant-in-Aid for JSPS Fellows. 

\bibliography{fe_JPSJ}

\begin{thebibliography}{10}
\bibitem{notation}
In this letter, $x$, $y$ axes are set to be directed along the nearest Fe-Fe
  bonds, in contrast to Refs.
  \cite{JPSJ.81.024604,JPSJ.80.073702,JPSJ.79.123707,SSC.152.701} where those
  are second nearest Fe-Fe bonds and the corresponding quadrupole is called as
  $O_{xy}$.
\bibitem{zdependence}
The same approximation has been used in Ref. \cite{PhysRevLett.107.137007} for
  the spin susceptibility which is found to well account for the inelastic
  neutron scattering experiment.
\end{thebibliography}
\end{document}